\documentclass[12pt]{article}
\usepackage{amsfonts}
\usepackage{amssymb}
\usepackage{amsfonts}
\usepackage{amsmath}
\usepackage{euscript}
\usepackage{fancyhdr}
\usepackage{slashed}
\usepackage{graphicx}
\usepackage{subfigure}
\usepackage{color}
\usepackage{hyperref}
\usepackage{float}
\usepackage[all]{hypcap}

\def\hybrid{\topmargin -20pt    \oddsidemargin 0pt
        \headheight 0pt \headsep 0pt
        \textwidth 6.25in       
        \textheight 9.25in       
        \marginparwidth .875in
        \parskip 5pt plus 1pt   \jot = 1.5ex}

\hybrid

\def\baselinestretch{1.2}

\catcode`\@=11

\def\marginnote#1{}
\newcount\hour
\newcount\minute
\newtoks\amorpm
\hour=\time\divide\hour by60
\minute=\time{\multiply\hour by60 \global\advance\minute by-\hour}
\edef\standardtime{{\ifnum\hour<12 \global\amorpm={am}%
        \else\global\amorpm={pm}\advance\hour by-12 \fi
        \ifnum\hour=0 \hour=12 \fi
        \number\hour:\ifnum\minute<10 0\fi\number\minute\the\amorpm}}
\edef\militarytime{\number\hour:\ifnum\minute<10 0\fi\number\minute}

\def\draftlabel#1{{\@bsphack\if@filesw {\let\thepage\relax
   \xdef\@gtempa{\write\@auxout{\string
      \newlabel{#1}{{\@currentlabel}{\thepage}}}}}\@gtempa
   \if@nobreak \ifvmode\nobreak\fi\fi\fi\@esphack}
        \gdef\@eqnlabel{#1}}
\def\@eqnlabel{}
\def\@vacuum{}
\def\draftmarginnote#1{\marginpar{\raggedright\scriptsize\tt#1}}

\def\draft{\oddsidemargin -.5truein
        \def\@oddfoot{\sl preliminary draft \hfil
        \rm\thepage\hfil\sl\today\quad\militarytime}
        \let\@evenfoot\@oddfoot \overfullrule 3pt
        \let\label=\draftlabel
        \let\marginnote=\draftmarginnote
   \def\@eqnnum{(\theequation)\rlap{\kern\marginparsep\tt\@eqnlabel}%
\global\let\@eqnlabel\@vacuum}  }

\def\preprint{\twocolumn\sloppy\flushbottom\parindent 2em
        \leftmargini 2em\leftmarginv .5em\leftmarginvi .5em
        \oddsidemargin -.5in    \evensidemargin -.5in
        \columnsep .4in \footheight 0pt
        \textwidth 10.in        \topmargin  -.4in
        \headheight 12pt \topskip .4in
        \textheight 6.9in \footskip 0pt
        \def\@oddhead{\thepage\hfil\addtocounter{page}{1}\thepage}
        \let\@evenhead\@oddhead \def\@oddfoot{} \def\@evenfoot{} }

\def\numberbysection{\@addtoreset{equation}{section}
        \def\theequation{\thesection.\arabic{equation}}}

\def\underline#1{\relax\ifmmode\@@underline#1\else
        $\@@underline{\hbox{#1}}$\relax\fi}

\def\titlepage{\@restonecolfalse\if@twocolumn\@restonecoltrue\onecolumn
     \else \newpage \fi \thispagestyle{empty}\c@page\z@
        \def\thefootnote{\fnsymbol{footnote}} }

\def\endtitlepage{\if@restonecol\twocolumn \else \newpage \fi
        \def\thefootnote{\arabic{footnote}}
        \setcounter{footnote}{0}}  

\catcode`@=12
\relax

\def\figcap{\section*{Figure Captions\markboth
        {FIGURECAPTIONS}{FIGURECAPTIONS}}\list
        {Figure \arabic{enumi}:\hfill}{\settowidth\labelwidth{Figure
999:}
        \leftmargin\labelwidth
        \advance\leftmargin\labelsep\usecounter{enumi}}}
 \relax
\def\tablecap{\section*{Table Captions\markboth
        {TABLECAPTIONS}{TABLECAPTIONS}}\list
        {Table \arabic{enumi}:\hfill}{\settowidth\labelwidth{Table
999:}
        \leftmargin\labelwidth
        \advance\leftmargin\labelsep\usecounter{enumi}}}
 \relax
\def\reflist{\section*{References\markboth
        {REFLIST}{REFLIST}}\list
        {[\arabic{enumi}]\hfill}{\settowidth\labelwidth{[999]}
        \leftmargin\labelwidth
        \advance\leftmargin\labelsep\usecounter{enumi}}}
 \relax
\makeatletter
\newcounter{pubctr}
\def\publist{\@ifnextchar[{\@publist}{\@@publist}}
\def\@publist[#1]{\list
        {[\arabic{pubctr}]\hfill}{\settowidth\labelwidth{[999]}
        \leftmargin\labelwidth
        \advance\leftmargin\labelsep
        \@nmbrlisttrue\def\@listctr{pubctr}
        \setcounter{pubctr}{#1}\addtocounter{pubctr}{-1}}}
\def\@@publist{\list
        {[\arabic{pubctr}]\hfill}{\settowidth\labelwidth{[999]}
        \leftmargin\labelwidth
        \advance\leftmargin\labelsep
        \@nmbrlisttrue\def\@listctr{pubctr}}}
 \relax
\makeatother
\newskip\humongous \humongous=0pt plus 1000pt minus 1000pt

\newif\ifdtup

\relax

\newcommand{\bse}{\begin{subequations}}
\newcommand{\ese}{\end{subequations}}
\def\be{\begin{equation}}
\def\ee{\end{equation}}
\def\ba{\begin{eqnarray}}
\def\ea{\end{eqnarray}}

\def\a{\alpha}

\def\g{\gamma}
\def\G{\Gamma}

\def\s{\sigma}

\def\cN{{\cal N}}

 \def\cN{{\cal N}} 
  
 \def\cT{{\cal T}}

\newcommand{\vev}[1]{{\left< {#1} \right>}}

\newcommand{\prt}[1]{{\left( {#1} \right)}}

\def\no{\noindent}

\def\IR{\relax{\rm I\kern-.18em R}}

\def\IR{\relax{\rm I\kern-.18em R}}
\def\IL{\relax{\rm I\kern-.18em L}}

\def\inv{^{\raise.15ex\hbox{${\scriptscriptstyle -}$}\kern-.05em 1}}

\def\bea{\begin{eqnarray}}
\def\eea{\end{eqnarray}}
\newcommand{\eq}[1]{(\ref{#1})}
\def\nn{\nonumber}

\newcommand{\la}[1]{\label{#1}}

\def\a{\alpha}      
       
\def\g{\gamma}  \def\G{\Gamma}

\def\s{\sigma}  
\def\t{\tau}

\definecolor{markcolor2}{rgb}{1,0,0}

\definecolor{markcolor3}{rgb}{0,1,0}

\begin{document}

\renewcommand{\theequation}{\thesection.\arabic{equation}}
\csname @addtoreset\endcsname{equation}{section}

\newcommand{\beq}{\begin{equation}}
\newcommand{\eeq}[1]{\label{#1}\end{equation}}
\newcommand{\ber}{\begin{eqnarray}}
\newcommand{\eer}[1]{\label{#1}\end{eqnarray}}
\newcommand{\eqn}[1]{(\ref{#1})}
\begin{titlepage}

\begin{center}


{\large
\bf The Imaginary Potential of Heavy Quarkonia Moving in Strongly Coupled Plasma}

\vskip 0.5in

{\bf M. Ali-Akbari$^{a}$, D. Giataganas$^{b,c}$, Z. Rezaei$^{d,e}$}
\vskip 0.4in
{\em
${}^a$  Department of Physics, Shahid Beheshti University G.C.,\\
Evin, Tehran 19839, Iran
\vskip .1in
${}^b$ Department of Nuclear and Particle Physics,\\
Faculty of Physics, University of Athens,\\
Athens 15784, Greece
\vskip .1in
${}^c$
National Institute for Theoretical Physics\\
School of Physics and Centre for Theoretical Physics\\
University of the Witwatersrand, Wits, 2050, South Africa
\vskip .15in
${}^d$Department of Physics, University of Tafresh, 39518-79611, Tafresh, Iran
\vskip 0.001in
${}^e$School of Particles and Accelerators, \\Institute for research in fundamental sciences (IPM),  19395-5531, Tehran, Iran
\\\vskip .1in
{\tt aliakbari@theory.ipm.ac.ir,\quad dgiataganas@phys.uoa.gr, \quad z.rezaei@aut.ac.ir}\\
}
\vskip .2in
\end{center}

\vskip .4in

\centerline{\bf Abstract}

\no

The melting of a heavy quark-antiquark bound state depends on the screening phenomena associated with the binding energy, as well as scattering phenomena associated with the imaginary part of the potential. We study the imaginary part of the static potential of heavy quarkonia moving in the  strongly coupled plasma. The imaginary potential dependence on the velocity of the traveling bound states is calculated. Non-zero velocity leads to increase of the absolute value of the imaginary potential. The enhancement is stronger when the quarkonia move orthogonal to the quark-gluon plasma maximizing the flux between the pair. Moreover, we estimate the thermal width of the moving bound state and find it enhanced compared to the static one. Our results imply that the moving quarkonia dissociate easier than the static ones in agreement with the expectations.
\end{titlepage}
\vfill
\eject


\noindent


\def\baselinestretch{1.2}
\baselineskip 19 pt
\noindent


\setcounter{equation}{0}

\tableofcontents

\section{Introduction}

Recently there has been a lot of interest in the heavy quarkonium suppression which has been observed in RHIC and LHC
\cite{Brambilla:2004wf,Brambilla:2010cs,Chatrchyan:2012lxa,Abelev:2012rv}.
The suppression is a signal of deconfinement and was suggested that the bound states dissociate in the hot thermal bath because of the color screening \cite{Matsui:1986dk}. There is a belief that the static potential apart from the usual real part develops an imaginary one, which contributes to the dissociation. For a bound state in the plasma, the imaginary part induced by the Landau-damping  of the gauge-fields that mediate interactions between the quarks, coming from the scattering of the gluons with the particles of the medium with momenta of order $T$ \cite{Laine:2006ns,Laine:2007gj,Beraudo:2007ky} and the quark-antiquark color singlet break up \cite{Brambilla:2008cx}. Around the deconfining temperature $T_c$,  it has been found by obtaining the Schr\"{o}dinger equation in a non-perturbative way, that the real part of the potential is modified milder while the imaginary part is growing. From lattice calculations it has been noticed that the increase of temperature which leads to increase of the number of collisions in the gluonic medium, plays important role in the destabilization of the heavy quarkonium, at least comparable to that of the screening effects \cite{Rothkopf:2011db,Laine:2007qy}. As a result, there is a possibility that the dominant mechanism for the quarkonium dissociation in particular scales, is due to the thermal width than the screening. Even more importantly, the screening and the Landau damping contribute simultaneously in the dissociation,  since in principle it is easier to dissociate a bound state with low binding energy  by gluon scattering phenomena, than a bound state with strong binding  energy. Therefore, the imaginary potential contribution in the dissociation of the bound state it is very interesting phenomenon to  be studied. In the bibliography there are several methods approaching the imaginary potential and extensive work has been already done to this direction \cite{Brambilla:2010cs,Brambilla:2010vq,Beraudo:2010tw}.

The width and the dissociation rate of a particular bound state depends on the temperature, the quark content and the quantum numbers. Most of the above mentioned weakly coupled results have been obtained using effective field theory (EFT) with an appropriate hierarchy of scales. For higher temperatures, where the heavy quark mass is the highest energy scale present, it can be integrated out to end up with non-relativistic QCD (NRQCD) \cite{Caswell:1985ui,Lepage:1992tx}. The study of the width of quarkonium states using NRQCD gives consistent results with the studies of the EFT calculation by appropriate fixing of the coupling constant \cite{Aarts:2011sm,Aarts:2010ek}.

Considering the fact that in the LHC, the heavy quarkonia are produced not only in large numbers but also with high momenta, an interesting expansion of these studies is the dependence of the thermal width on one more parameter, the speed of the bound state. In this paper we study the effects of the velocity on the imaginary potential and the thermal width of the heavy bound state in the strong coupling limit using gauge/gravity duality methods.

It is expected that there is a dependence of the decay rate on the speed. This is confirmed for the bottomonium S wave correlation in \cite{Aarts:2012ka}. Similar studies were performed in \cite{Escobedo:2013tca} using the EFT for temperatures much smaller than the heavy quark mass, where the dependence of the thermal width on the velocity was also noticed. In this work, different hierarchy of scales lead to a different qualitative dependence of the decay width on the velocity. Moreover, using the perturbative QCD and several approximations, the thermal width was found to have comparable contributions from the leading order (LO) and the next to the leading order (NLO) due to the fact that the gluo-dissociation diagrams that contribute in the LO, are suppressed in the NLO. Calculating the total contribution, it has been found that the increase of speed leads to higher decay rate for the quarkonia, and therefore to higher suppression rates \cite{Song:2007gm}.

An alternative powerful tool for calculating observables in the strong coupling limit is the AdS/CFT duality \cite{Maldacena:1998re,Witten:1998qj}. There is an extended methodology for the calculation of several observables in the isotropic QGP at equilibrium \cite{CasalderreySolana:2011us} as well as in anisotropic ones \cite{Giataganas:2012zy,Giataganas:2013lga}. In the context of the gauge/gravity dualities there are different ways that lead to a complex static potential, e.g. \cite{Albacete:2008dz,Noronha:2009da,Hayata:2012rw}. Although there are advantages and disadvantages of each way, a more complete picture of the imaginary potential generation technique in the AdS/CFT duality is still needed. Some of the methods use a modified way for the UV divergences cancellation. Taking into account however that there are physical and mathematical reasons to follow the known cancellation schemes of UV divergences in the Wilson loop minimal surfaces \cite{Maldacena:1998im,Drukker:1999zq,Chu:2008xg}, a better arguing is needed when the Wilson loop UV divergences cancellation is done with alternative methods. In our case we use the usual divergence cancellation technique by subtracting the infinite masses of the quarks and choose an approximate method to generate the imaginary potential from the fluctuations at the deepest point in the bulk of the string world-sheet \cite{Noronha:2009da}. It has been found that for a static Q\={Q} pair the imaginary potential can be calculated for any background with diagonal metric through a simple formula derived in \cite{Giataganas:2013lga,Giataganasimv}. Using this method the imaginary part and the thermal width was calculated in the axion deformed spatially anisotropic quark gluon plasma and its dependence on the anisotropic parameter was found \cite{Giataganasimv}. Moreover, an estimate of how the thermal width changes with the shear viscosity to entropy density ratio was given in \cite{Finazzo:2013rqy} and a study of the higher order corrections was done in \cite{Fadafan:2013coa}.

Apart from the static pair in the gauge/gravity dualities, there is a considerable amount of work done studying the dissociation of moving bound states \cite{Liu:2006he,Liu:2006nn,Caceres:2006ta,Chernicoff:2006yp, Avramis:2006em,
Chernicoff:2006hi,Peeters:2006iu,Liu:2008tz}. The usual trick is to boost the pair into a frame which is at rest while the hot wind of quark-gluon plasma (QGP)  moves against it. Calculating  the expectation value of the Wilson loop in this setup, the dependence of the binding energy on the velocity was found. The screening length of the bound state in an isotropic plasma, depends on the angle between the direction of the Q\={Q} and the velocity of the wind. Increasing the angle from zero (parallel motion) to $\pi/2$   (perpendicular motion) the screening length for a fixed velocity is decreasing. The screening length turns out to be decreasing with increasing velocity and being proportional to the $\prt{energy~density}^{-1/4}$. The velocity scaling in the screening length function has been shown to be a universal feature for several strongly coupled theories. However this particular power dependence is modified in anisotropic gauge/gravity dualities \cite{Chernicoff:2012bu}. This is not completely unexpected since when spatial anisotropies are present, several universal relations are violated like inequalities between the components of the Langevin coefficients \cite{Giataganas:2013hwa,Giataganas:2013zaa} and the known shear viscosity over entropy ratio \cite{Rebhan:2011vd}.

In this paper, employing gauge/gravity duality techniques we calculate the imaginary potential for the two extreme pair alignments, one transverse and one parallel to the speed of the wind, and we estimate the thermal width. We find that the imaginary potential does depend on the velocity of the moving pair as well as the angle between the alignment of the pair and the direction of velocity. Increase of speed, independently of the angle of motion, leads to increase of the absolute value of the imaginary potential. It leads also to decrease of the minimum pair distance $L$ that the $ImV$ takes the non-zero value. Using certain approximations we find that the thermal width is a monotonically increasing function of the speed.

Increasing the angle for a fixed velocity, we find that the absolute value of the imaginary potential and the thermal width are increased. This can be explained by the fact that the flux of the QGP between the quark-antiquark pair increases and becomes maximum for perpendicular motion, allowing the pair to dissociate earlier. Therefore, using the gauge/gravity duality we find that the screening phenomena and the Landau damping phenomena, both contributing to the dissociation of the Q\={Q} pair in the same qualitative way while  depending on the angle and the velocity of the motion. Notice that the screening and the Landau damping operate simultaneously in the dissociation,  since in principle is easier to dissociate a bound state with low binding energy by scattering than a bound state with strong binding.

This paper has the following structure. In section 2 we use a generic boosted metric with non-diagonal elements and obtain the imaginary potential and other useful formulas in terms of the generic background metric elements. The work is done for both the transverse and parallel directions of motion of the pair. In section 3 we apply our analysis to the
gravity dual of the finite temperature $\cN=4$ sYM and we find the dependence of the $ImV$ on the angle and the velocity of motion. We analyze the effects of the velocity and the relative angle of the motion of the pair, on the imaginary potential and estimate the thermal width. In section 4, we compare our results to others known in the literature and we point out the similarities. In the final section we summarize and discuss our findings.

\section{Setup for the moving Quarkonium in Finite Temperature}

In this section we present a generic analysis of the imaginary part of the static potential for a quark anti-quark pair moving in the plasma. The gravity background has the following form
\be\label{background} %
ds^{2}=-\tilde{G}_{00}dx_0^{2}+\sum_{i=1}^{3}\tilde{G}_{ii}dx_i^2+
\tilde{G}_{uu}du^2~,
\ee %
where all the metric elements depend on the radial distance $u$ only, and the rest of the directions are isometries of the space.
Boosting this metric along the $x_3$ direction with $t=\g\prt{t'-v x'_3}~, x_3=\g\prt{-v t'+x'_3}$ and dropping the primes we get to %
\be\label{background-boost} %
ds^{2}=G_{00}dx_0^{2}+\sum_{i=1}^{2}G_{ii}dx_i^2+G_{33}dx_3^2+
2G_{03}dx_0 dx_3+G_{uu}du^2,
\ee %
where the metric components in terms
 of the original metric (\ref{background}), are
\bea\label{metric-general}%
&&G_{00}=\gamma^2(\tilde{G}_{00}+v^2 \tilde{G}_{33})~,\quad G_{uu}=\tilde{G}_{uu}~, \quad  G_{i i}=\tilde{G}_{i i},\\
&&G_{33}=\gamma^2(\tilde{G}_{00}v^2+\tilde{G}_{33}) , \quad
 G_{03}=-\gamma^2 v (\tilde{G}_{00}+\tilde{G}_{33})~,
\eea %
where $\gamma^2=1/(1-v^2)$.
We consider the usual rectangular Wilson loop with a short side of length $L$ in the spatial direction and a long side along a time direction. We align the loop in two different ways, one where the short side of the Wilson loop is parallel to the direction of the wind and one transverse to it
\bea
&&x_3=\s~,\quad x_0=\t~,\quad u=u\prt{\s}~, \quad \mbox{ parallel to the wind,}\\
&&x_1=\s~,\quad x_0=\t~, \quad u=u\prt{\s}~,\quad \mbox{ transverse to the wind.}
\eea
Using the Nambu-Goto action and working generally we obtain
\be\label{action-general} %
S_{NG}=\frac{{\cal{T}}}{2\pi\alpha'}\int_{-L/2}^{L/2} d\sigma\sqrt{g(u)u'^2+F(u)},
\ee %
where we have defined
\bea\label{Mu} %
&&g(u):=-G_{00} G_{uu}~,\quad f_i\prt{u}:=-G_{00} G_{ii}~,\\
&&\label{Nu}%
F(u):=\left\{%
\begin{array}{ll} %
    G_{03}^2+f_3,\quad \parallel\mbox{to the wind}~, \\
    f_1,\quad\quad\quad\quad \perp\mbox{to the wind.} \\
\end{array}%
\right.
\eea %
The Hamiltonian is a constant of motion equal to $-c$ and is obtained as
\be \label{hamiltonian} %
c=\frac{F(u)}{\sqrt{g(u)u'^2+F(u)}}~,%
\ee %
while the turning point $u_0$ of the string world-sheet can be found by solving the equation $c^2=F(u_0)$. The distance $L$ between the quark and anti-quark pair is given by
\bea\label{distance-general} %
L=2 \int_{u_0}^{\infty} du
\left[\frac{F}{g}\left(\frac{F}{F_0}-1\right)\right]^{-1/2}~,
\eea %
where $F_0:= F(u_0)$. To extract $Im V_{Q\bar{Q}}$ we consider
thermal worldsheet fluctuations about the classical configuration. 
The profile of the string configuration is a string with a turning point at $u=u_0$ at $\s=0$. The expansion around this point is done by considering the fluctuations is $u(\sigma)\simeq u_{0}+\sigma ^2 u''_{0}/2$.
One can separate the real and imaginary part of the action with the imaginary contributions coming from the region around the turning point of the string, defined by the $\s_c$ as follows
\be\la{actione1}
iS\propto\frac{-\cT}{2 \pi \a'}\prt{\int_{|\s|<\s_c} d\s \sqrt{-c_2-\s^2 c_1}+i \int_{|\s|>\s_c}d\s \sqrt{F\prt{u}+g\prt{u} u'^2}}~,
\ee
where
\bea\label{c1c2}%
c_1\simeq  \frac{1}{2} u''_{0}(2g_0 u''_{0}+F'_{0})~,\qquad
c_2\simeq  F_{0}+\delta u F'_{0}+\frac{1}{2}\delta u^2
F''_{0}~.
\eea
From \eq{hamiltonian} we find that $u''_{0}=F'_{0}/\prt{2g_0}$. To obtain imaginary part in the action we need $c_2<0$  and $\s$ to be between the roots of the quadratic equation, then $\s_c=\sqrt{-c_2/c_1}$. Using equation \eq{actione1} and \eq{c1c2} we obtain the expression of the imaginary potential
\be\label{ImV}
{\rm{Im}}V_{Q\bar{Q}}=\frac{\sqrt{g_0}}{2\sqrt{2}\a'}\prt{\frac{F_0}{|F_0'|}-\frac{|F_0'|}{2 F_0''}}~.
\ee
Notice that this formula has similar functional form to the formula derived in \cite{Giataganas:2013lga,Giataganasimv} for the imaginary part of the potential of a static Q\={Q} pair. However the functions $F$ have different structure in the two cases and here are given by \eq{Nu}.

From the imaginary potential one can in principle extract the thermal width as
\be\label{thermalwidth1o}%
\Gamma_{Q\bar{Q}}=-\langle \psi|{\rm
Im}V|\psi\rangle~,
\ee%
where $|\psi\rangle$ can be thought as the ground state of the unperturbed static Coulomb potential.

In the next sections we apply our formulas to a Q\={Q} pair moving in the
boosted  finite temperature $AdS_5$ metric  and we study the effect of the motion on imaginary potential of a heavy quark pair moving in the dual $\cN=4$ sYM plasma.

\subsection{Imaginary Potential of a Moving Pair in Finite Temperature}

The boosted finite temperature $AdS$ metric along the $x_3$ direction reads
\bea\nn
ds^2= &&\left(\frac{u}{R}\right)^{2}\prt{ -\left(1-\gamma^2\left(\frac{u_h}{u}\right)^{4}\right) dt^2+d\textbf{x}^2 +
\left(1+\gamma^2 v^2\left(\frac{u_h}{u}\right)^{4}\right) dx_3^2} \\
&&+2\gamma^2 v\left(\frac{u_h^2}{R u}\right)^{2}dt d x_3+ \frac{du^2}{\left(\frac{u}{R}\right)^{2}\left(1-\left(\frac{u_h}{u}\right)^{4}\right)}~,
\eea %
where $R$ is the $AdS_5$ radius and we take it equal to unit. The functions $g(u)$ and $F(u)$ can be read from \eq{Mu} and \eq{Nu} as
\be %
g(u)=\frac{u^{4}-\gamma^2 u_h^{4} }{u^{4}-u_h^{4}}
\ee %
and
\bea
F(u)=\left\{%
\begin{array}{ll} %
    u^{4}-u_h^{4} ,\qquad \quad \parallel\mbox{ to wind,} \\
    u^{4}-\gamma^2 u_h^{4} ,\qquad \perp\mbox{ to wind.}\\
\end{array}%
\right.
\eea %
Applying our results of the previous section to this particular metric, we get for the two possible directions of motion.

\subsubsection{Pair alignment parallel to the wind}

The turning point of the string depends on the direction of the motion and is given by $u_{0}^{4}=c^2 +u_h^{4}~.$ The distance between quark and anti-quark can be found using  the \eq{distance-general} and it is given by
\bea\la{ll1}
\frac{L}{2}=
   \;\int_{u_0}^\infty du
\sqrt{\dfrac{(u_0^4-u_h^4)(u^4-\gamma^2 u_h^4)}{(u^4-u_h^4)^2(u^4-u^4_{0})}}~.
\eea %
In order to have real values for the length we need to satisfy $u_0^{4}\geq u_h^{4}\gamma^2$.  This integral can be integrated to give elliptic functions which can be approximated for small lengths to polynomials. The imaginary part of the potential can be written analytically in terms of the radial coordinate $u$.  Using \eqref{ImV} we get
\bea\label{ImV-d31}
{\rm Im} V_{Q\bar{Q}}=
   -\dfrac{1}{8\sqrt{2}
\alpha'}\dfrac{\sqrt{u_{0}^4-u_h^4\gamma^2}}{u_{0}^3\sqrt{u_{0}^4-u_h^4}}\left(
u_h^4-\dfrac{u_{0}^4}{3}\right)~.
\eea %
We are interested in studying the bound states for which the $Im V<0$.
This condition constrains the turning point and by taking into account the previous constraint for real length we get
\be\la{contra1}
\gamma^2 u_h^4<u^4_{0}<3u_h^4~.
\ee
This condition puts directly an upper limit on the velocity  $v<\sqrt{2/3}$, in order to have negative $ImV$ for motion along the direction of the wind.

\subsubsection{Pair alignment transverse to the wind}

For the pair alignment transverse to the wind the turning point of the string  depends additionally on the velocity of the wind and is given by $u_{0}^{4}=   c^2 +\gamma^2 u_h^{4}$, where again we get $u_0^{4}\geq u_h^{4}\gamma^2$. The distance between quark and anti-quark takes the form
\bea\label{ll2}%
\frac{L}{2}=
    \;\int_{u_0}^\infty du
\sqrt{\dfrac{u_{0}^{4}-u_h^{4}\gamma^2}{(u^4-u_h^4)(u^4-u^4_{0})}}~,
\eea %
which when integrated leads again to elliptic functions and at particular limits can be approximated to a polynomial form.
The imaginary part of the potential can be written analytically in terms of the radial coordinate $u$ and has an additional factor $\g$ compared to the parallel case
\bea\label{ImV-d32}
{\rm Im} V_{Q\bar{Q}}=
   -\dfrac{1}{8\sqrt{2}
\alpha'}\dfrac{\sqrt{u_{0}^4-u_h^4\gamma^2}}{u_{0}^3\sqrt{u_{0}^4-u_h^4}}\left(
\gamma^2 u_h^4-\dfrac{u_{0}^4}{3}\right)~ .
\eea %
To have negative $ImV$ we are interested in string world-sheets that satisfy the following constraint for the turning point $u_0$
\bea\label{contra2}
   \gamma^2 u_h^4<u^4_{0}<3\gamma^2 u_h^4~.
\eea %
Notice that the left inequality is trivially satisfied from the turning point equation, while the right inequality does not set a bound to  the speed of the pair directly, in contrast to what happens for the parallel motion to the wind \eq{contra1}. Alternatively this inequality can be seen as setting a constraint on the proximity of the turning point of the world sheet to the horizon of the black hole for a given speed.

\section{Analysis of the Imaginary Potential}

In this section we study  the behavior of the imaginary potential with respect the velocity of the wind, the size of the bound state and the direction of motion of the pair. We treat separately the parallel and transverse cases.

\subsection{Bound State Parallel to the Wind}

The bound states we study are those that have negative imaginary potential, real values of the length and to be in the energetically favorable and stable side of the double valued function $ReV\prt{L}$ .
The first two requirements are satisfied by using the constraint \eq{contra1} and the other is imposed in the analysis.

\begin{figure}
\begin{minipage}[ht]{0.5\textwidth}
\begin{flushleft}
\centerline{\includegraphics[width=70mm]{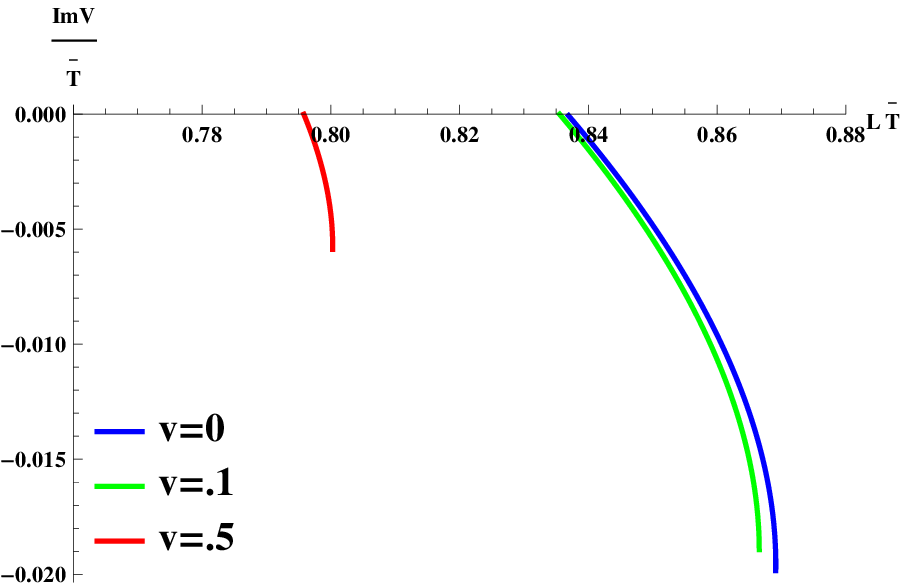}}
\caption{\small{Dependence of the imaginary potential on the distance between the pair for parallel motion and fixed
velocities $v=0,~0.1,~0.5$, corresponding to the curves from right to left respectively.  The imaginary part is generated for $L_{min}$ and decreases until the value  $L_{max}$. Increase of the speed causes the imaginary potential to be generated for smaller distances between the pair, implying easier dissociation.  The dimensionless quantities are constructed with the use of the temperature scaled as $\bar{T}:=\pi T$.
\vspace{0cm}}}\label{imvp22}
\end{flushleft}
\end{minipage}
\hspace{0.3cm}
\begin{minipage}{0.5\textwidth}
\begin{flushleft}
\centerline{\includegraphics[width=70mm]{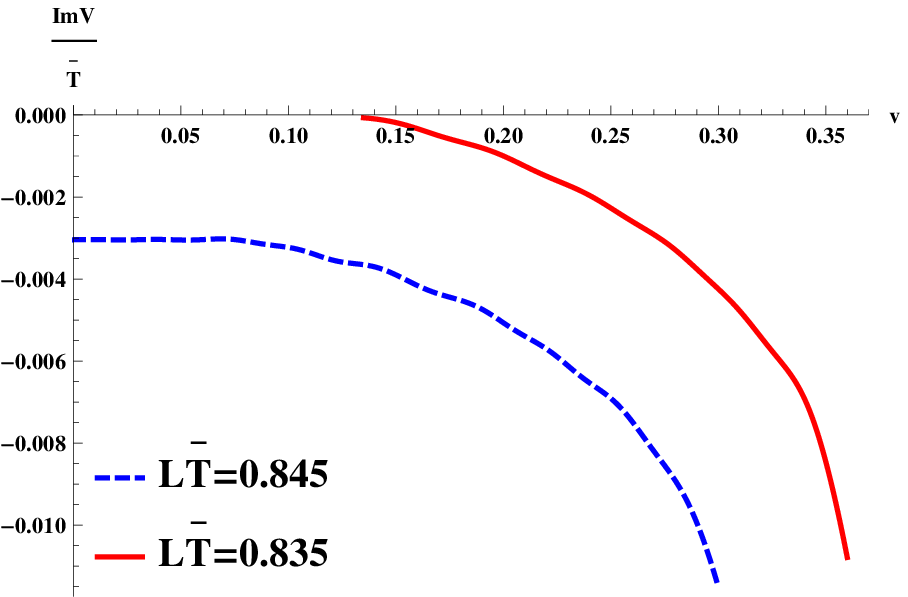}}
\caption{\small{Dependence of the $ImV$ on the velocity of the pair, for two fixed lengths $L$ and parallel motion to the wind. As the speed increases the imaginary potential is decreasing with a higher rate. Bound states of larger sizes have lower imaginary potential. Notice that for the lower size of the bound state, the imaginary potential takes zero value for non-zero speed. \label{imvpl2}
\vspace{2.cm}}}
\end{flushleft}
\end{minipage}
\end{figure}

Using the expression  \eq{ll1} and \eq{ImV-d31} we find numerically the relations $ImV\prt{L}$ for fixed values of the velocity and $ImV\prt{v}$ for fixed distances between the pair. We find that for fixed velocities and increasing size of the bound state, the absolute value of $ImV$ increases. This later behavior generalizes the findings of the zero velocity \cite{Noronha:2009da,Giataganasimv} to non-zeroth speeds. Moreover, the imaginary part of the potential is generated at lower distances between the pairs. It is also interesting to notice that there is a maximum value for the absolute value of $ImV$, similar to a turning point of the real potential and as the velocity increases, it decreases. This peculiarity exists also for zero velocities, and a reason that appears is the form of the function $L\prt{u_0,u_h,v}$ for the distance between the pair, which is the same with binding energy analysis. This was discussed in more detail in \cite{Giataganasimv} and a straightforward proposed way to deal with it for the calculation of the thermal width, is an extrapolation of the straight part of the curve to larger lengths. The findings described in this paragraph are shown in Figure \ref{imvp22}.

Fixing the size of the bound state and changing the velocity we find that increase of the velocity leads to increase of the absolute value of the imaginary potential. This implies that the pair moving parallel to the wind is more likely to decay as its velocity is increased. Higher speeds lead to higher rate of increase in the $|ImV|$ and also the generation of it for shorter lengths $L$ between the pair. We also notice  that larger distances $L$ have always larger values of $|ImV|$ for any fixed velocity. These findings are depicted in the Figure \ref{imvpl2}.

\begin{figure}
\begin{minipage}[ht]{0.5\textwidth}
\begin{flushleft}
\centerline{\includegraphics[width=70mm]{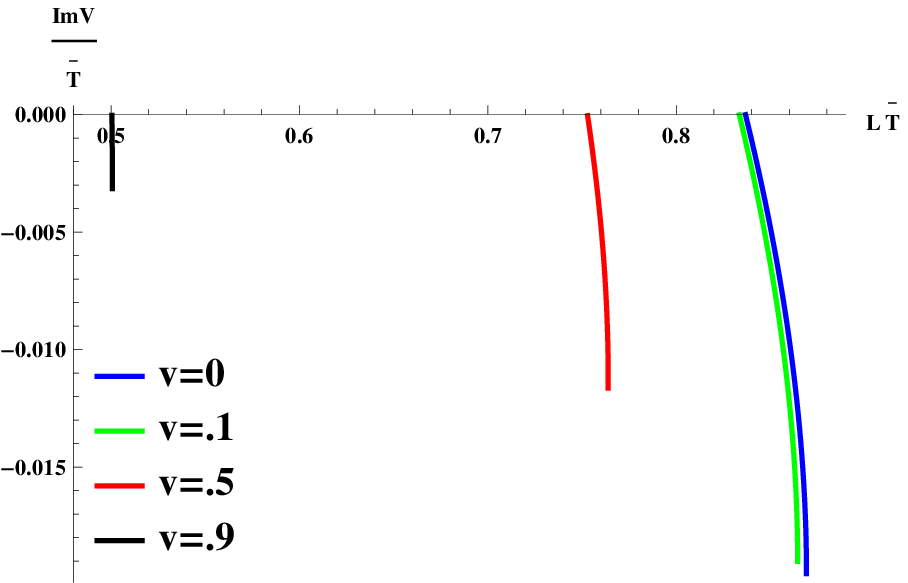}}
\caption{\small{Dependence of the $ImV$ on the distance between the pair for transverse motion for fixed velocities $v=0,~0.1,~0.5,~0.9$ corresponding to the curves from right to left respectively.  Notice that in the transverse case there is a solution for $v=0.9$ with negative $ImV$ that did not exist in the case of parallel motion.  The effect of the velocity in the transverse motion is much stronger than the parallel motion.
\vspace{0cm}}}\label{imvt011}
\end{flushleft}
\end{minipage}
\hspace{0.3cm}
\begin{minipage}{0.5\textwidth}
\begin{flushleft}
\centerline{\includegraphics[width=70mm]{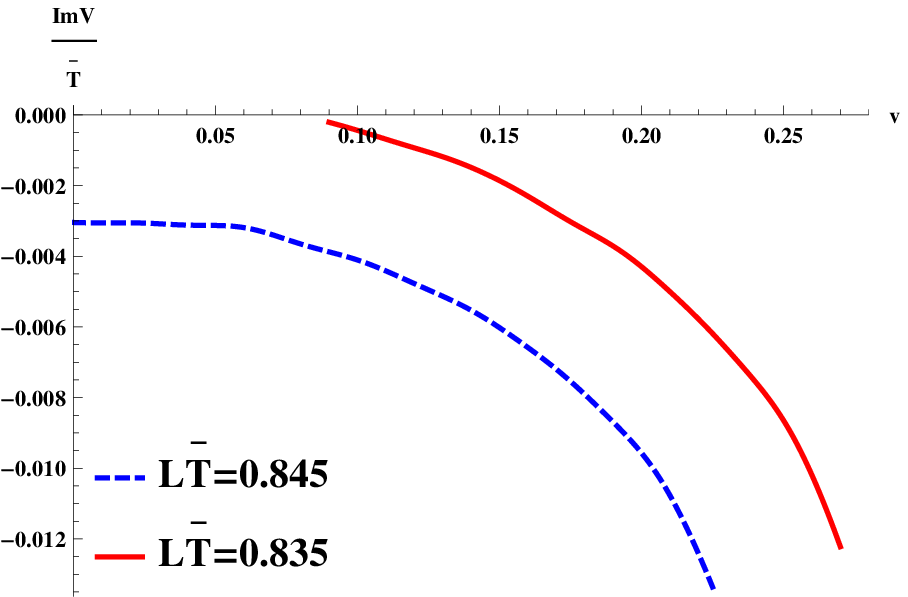}}
\caption{\small{Dependence of the imaginary potential on the velocity of the pair for two fixed lengths $L$, for the transverse motion. As the speed increasing the imaginary potential decreases, while quarks in larger distances have lower imaginary potential for all the speeds. Qualitatively this is the same behavior noticed for the parallel motion. \label{imvtl1}
\vspace{.5cm}}}
\end{flushleft}
\end{minipage}
\end{figure}

\subsection{Bound State Transverse to the Wind}

For a pair moving orthogonal to the wind we use the equations \eq{ll2} and \eq{ImV-d32} to find numerically the functions  $ImV\prt{L}$  and $ImV\prt{v}$, for fixed values of the velocity and for fixed distances between the pair respectively. Qualitatively the results are similar to the parallel case although quantitatively the results for the transverse motion of the pair are more sensitive to the speed.

Like the parallel case we find that increase of length for constant velocity leads to increase of the $|ImV|$. Here however we have solution with a negative imaginary part for velocities $v=0.9$ in contrast to the parallel case. We also notice that the $ImV$ is decreasing faster than  the parallel case. These findings are depicted in Figure \ref{imvt011}. By fixing the lengths of the pair and changing the velocities we see that increase of velocity leads to increase of the absolute value of the imaginary potential (Figure \ref{imvtl1}).

\subsection{Transverse vs Parallel Motion}

\begin{figure*}[!ht]
\begin{minipage}[ht]{0.5\textwidth}
\begin{flushleft}
\centerline{\includegraphics[width=70mm]{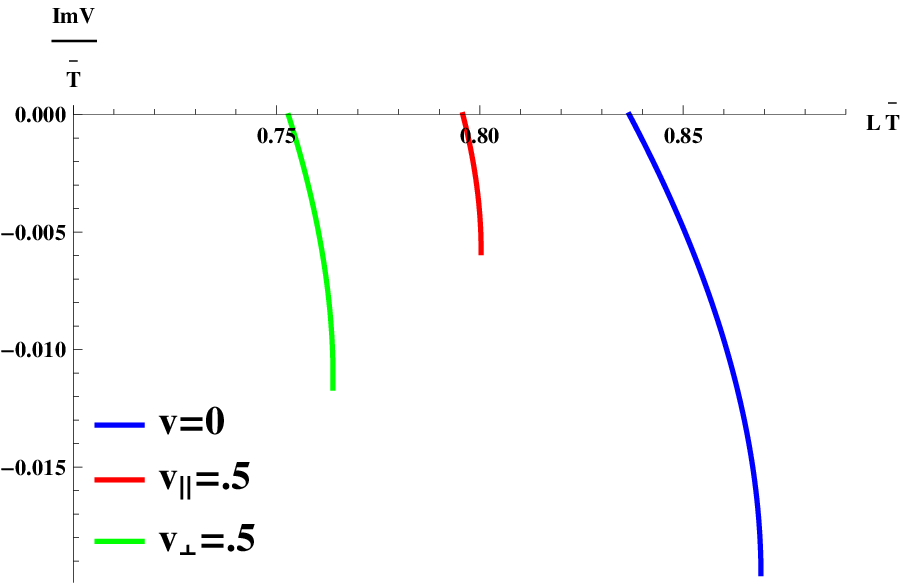}}
\caption{\small{Dependence of the imaginary potential on the distance of the pair for different directions of motion. The velocities $v=0,~v_\parallel=0.5,~v_\perp=0.5$ correspond to the curves from right to left respectively. Orthogonal motion results in a reduced imaginary potential compared to the parallel case, implying easier disassociation of the heavy bound state.
\vspace{0cm}}}\label{imvcom1}
\end{flushleft}
\end{minipage}
\hspace{0.3cm}
\begin{minipage}{0.5\textwidth}
\begin{flushleft}
\centerline{\includegraphics[width=70mm]{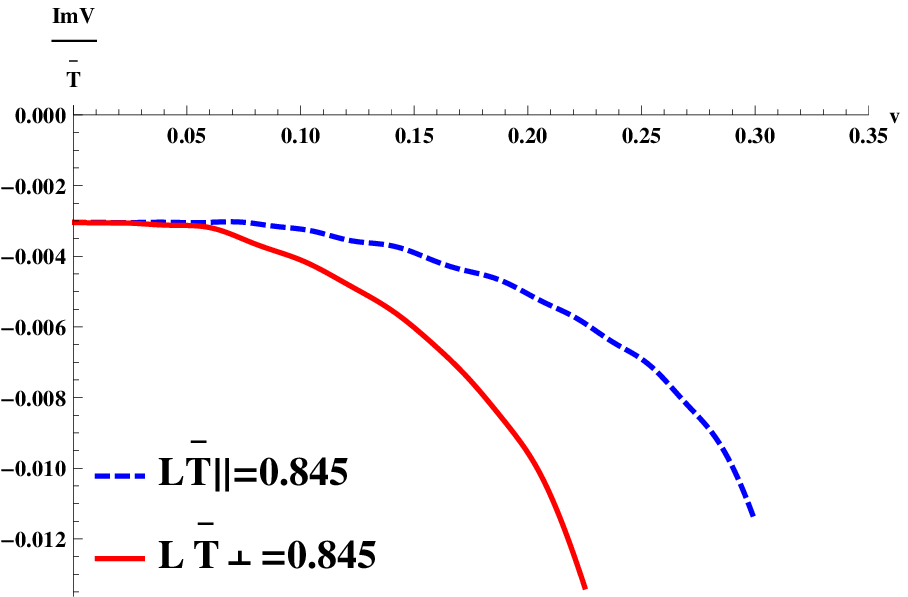}}
\caption{\small{ The imaginary potential depending on the velocity of the pair for two fixed lengths $L$ between the pair, compared for parallel and transverse motion to the wind. As the speed increases, the reduction of the $ImV$ for the transverse motion is stronger compared to the parallel motion to the wind. \label{imvtplcom}
\vspace{.5cm}}}
\end{flushleft}
\end{minipage}
\end{figure*}

It is very interesting to compare the effect of the velocity on the imaginary potential for a pair orthogonal to the wind and a pair moving parallel to the wind. We  notice that the imaginary potential for the motion transverse to the wind is modified stronger than the parallel case. The particular behavior can be observed in Figure \ref{imvcom1}. By placing a pair of the same size moving in a QGP, we observe that for low speeds the $ImV$ values are very close for both directions of motion. As the velocity increases the imaginary potential appears to reduce stronger for the motion transverse to the wind in comparison with the parallel motion, implying a higher decay rate for the transverse motion (Figure \ref{imvtplcom}). This is physically expected since the pair is easier to dissociate when oriented orthogonal to the direction of the wind where the flux of the QGP between the quark-antiquark pair is maximum for perpendicular motion.

Notice that the implication of the dissociation mechanism corresponding to the imaginary potential matches qualitatively the screening effects results. The binding energy analysis shows that the screening length is minimum for motion transverse to the wind, and for motion in any direction it is reduced compared to the static case \cite{Liu:2006he}. Therefore, the imaginary part of the potential contributes in the same way to the dissociation as the binding energy.

\subsection{Thermal Width}

The thermal width can be found from the imaginary potential. Following one of the suggestions of \cite{Giataganasimv}, we use an approximation where the almost straight $ImV\prt{L}$ curve before the turning point, is extended to larger values of $L$. To find the thermal width we choose a low speed and  isolate the $1/L$ dependence of the binding energy which depends on the velocity and subsequently we specify the velocity dependent "Bohr radius" of the wave-function. Having therefore specified the wavefunction appearing in \eq{thermalwidth1o} we integrate the corresponding equation using the extrapolated straight line $ImV\prt{L}$, where we choose as the minimum length the point where the $ImV$ becomes non-zero and integrate to an infinite maximum.

\begin{figure*}[!ht]
\begin{minipage}[ht]{0.5\textwidth}
\begin{flushleft}
\centerline{\includegraphics[width=70mm]{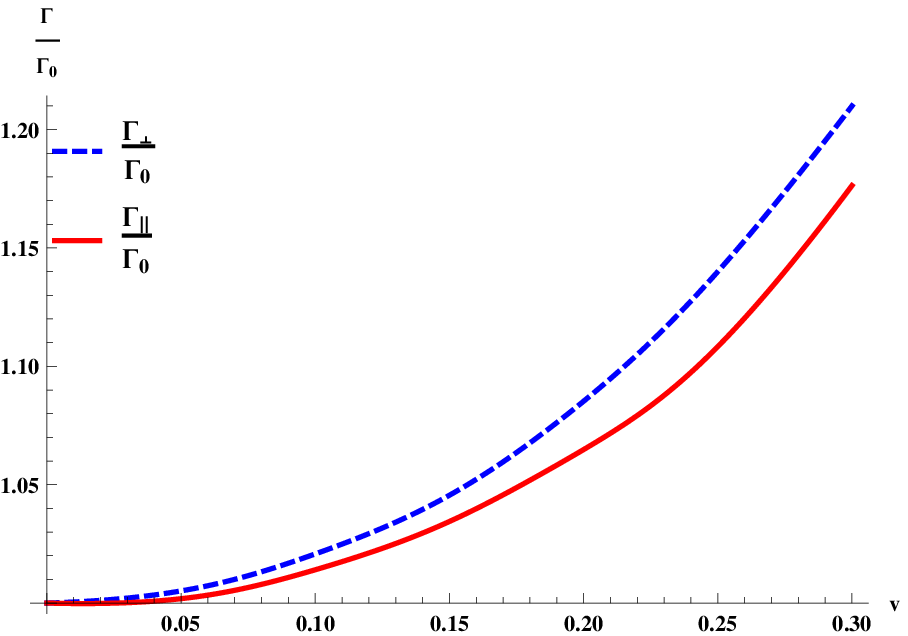}}
\caption{\small{The estimated thermal widths normalized with the zero velocity thermal width. Both ratios are larger than the unit indicating the increase of the thermal width for a moving bound state.
\vspace{0cm}}}\label{fig:gtr1}
\end{flushleft}
\end{minipage}
\hspace{0.3cm}
\begin{minipage}{0.5\textwidth}
\begin{flushleft}
\centerline{\includegraphics[width=70mm]{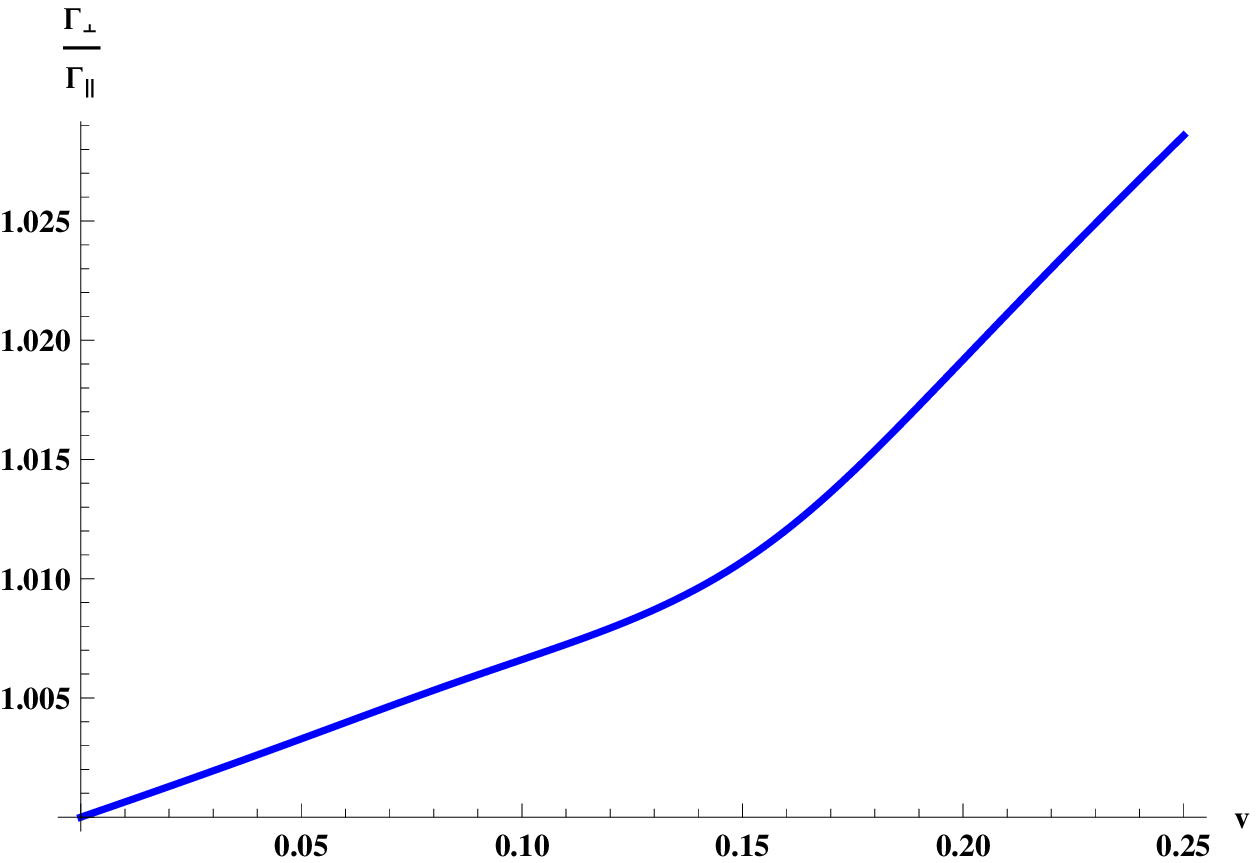}}
\caption{\small{ The ratio of the thermal widths for transverse and parallel motion to the wind. Notice the slowly increasing ratio with the speed of motion. \label{fig:gtr2}
\vspace{.3cm}}}
\end{flushleft}
\end{minipage}
\end{figure*}

For low speeds we get the following inequalities for the ratios
\be
\frac{\G_\perp}{\G_0}>\frac{\G_\parallel}{\G_0}>\frac{\G_\perp}{\G_\parallel}>1~,
\ee
where $\G_0,~\G_\parallel,~\G_\perp$ are the thermal widths for a static pair, a pair moving parallel to the wind and a pair moving transverse to it respectively. The pair moving orthogonal to the wind has increased thermal width compared to that of the moving pair aligned parallel to the wind as expected. Moreover, the thermal width for a moving pair is always increased compared to a static pair irrelevant of the direction of motion. Notice, that in our  comparison between the moving pair and the static one, we suppose that there are no other parameters except the ones mentioned above, that depend on the velocity and affect the result. We  present our results for low speeds in the Figures \ref{fig:gtr1} and \ref{fig:gtr2}.

Therefore, in agreement with what is naturally expected we find that increase of speed leads to increase  of the decay rate of a pair. As the moving pair rotates from a parallel motion to larger angles to get  eventually aligned to a transverse motion, such that the flux of the QGP
that flows between the pair becomes maximum, the imaginary potential and the thermal width are increased.

\section{Qualitative comparison with other methods and results}

In this section we attempt to discuss qualitatively and compare our results with the ones obtained in weakly coupled or strongly coupled plasmas using alternative methods. Our purpose here is to examine if there is a consistency of the known results and to point out possible similarities and differences. We remark that although this discussion is very interesting, part of it refers to results obtained at different ranges of coupling constant, between theories with different degrees of freedom or field content and therefore there is no reason a priori for the findings to agree. Nevertheless, we find that there is an extensive qualitative agreement between our results and other results in bibliography.

The dependence of our results on the temperature of the plasma for a fixed velocity is qualitatively the same to the one obtained in \cite{Noronha:2009da,Giataganasimv}. Increase of the temperature leads to increase of the imaginary potential, for a fixed velocity. In \cite{Escobedo:2013tca} the weakly coupled quarkonium propagating through a quark-gluon plasma has been studied. The results obtained there depend on the hierarchy of scales involved in the calculation: the binding energy, the size of the bound state, the screening mass and the heavy mass of the quark. In relatively low temperatures $TL\ll1$ a decreasing decay width as a function of the velocity was found, while in higher temperatures $TL\gg 1$ the width increases for low velocities and decrease for ultra-relativistic ones.

Recent lattice QCD computations with two flavors of light quarks using non-relativistic dynamics and at relatively low temperatures, show no dependence or a weak dependence of the thermal width on the velocity that is within the computational error \cite{Aarts:2012ka}. This is not in contrast to what we find here since the dependence of the $ImV$ and the thermal width on the velocity are continuous smooth functions that for low velocities are expected to give small contributions. The thermal width of heavy quarkonia has been also investigated in the perturbative QCD.  The NLO contribution to the decay width was found to be larger than the LO due to suppression of the gluo-dissociation diagrams at that order \cite{Song:2007gm}. The sum of the LO and the NLO lead to an increase of thermal width for increasing velocity and this is in agreement with what we have found. We remark that there are certain approximations in this work, since the same formula is used for the parton scattering and the gluo-dissociation widths and the Pauli blocking is ignored \cite{Brambilla:2013dpa}.


\section{Conclusions}

In this paper we have extended the studies \cite{Noronha:2009da, Giataganasimv} of the imaginary potential for a moving pair of quarks in Quark-Gluon Plasma. We have worked with a boosted metric with non-diagonal terms and we have considered a moving string with endpoints on the boundary and a turning point close to the horizon. We have analyzed  two extremal directions of motion, one that the pair is aligned parallel to the direction of speed of the wind and another that the pair is aligned transverse to the direction of the wind such that the flux of the plasma between the pair is maximized. Fluctuations around the turning point lead to an imaginary part of the potential which is given by a formula that depends on the velocity and turns out that it can be written in a functional form as a generalization of the corresponding formula derived in \cite{Giataganas:2013lga, Giataganasimv}.

For moving pair we find that the imaginary potential is increased in absolute value compared to that of a static pair with the same length. Increase of speed leads to further increase of the $ImV$ in both directions. Furthermore, increase of length leads to enhanced  $ImV$
as has been also noticed for the static pair. A further interesting result is that the pair moving orthogonal to the wind is affected stronger compared with the one moving parallel to it, leading to a stronger modification of the $|ImV|$. This can be compared with the binding energy analysis, where the screening length decreases as the pair is rotated to transverse angle with respect to the wind \cite{Liu:2006he}. The imaginary potential and thermal width analysis gives results to that direction too, since the $|ImV|$ is increased. This finding is natural, since when the angle between the direction of the pair and the velocity is increased, the flux of the plasma that passes between the quark-antiquark pair is increased and becomes maximum when the pair is aligned orthogonal to its speed. This is an indication that the pair is expected to be disassociated easier as the angle increases. 

To study the ratio of the thermal width at low speeds we are making several reasonable approximations. We isolate the Coulombic part of the binding energy to obtain a Coulombic wavefunction which depends on the speed and we calculate the quantity $\vev{\psi|ImV|\psi}$. At low velocities we find that the thermal width is increased stronger for motion in the orthogonal direction compared to the parallel one. Moreover, non-zero Q\={Q} velocities lead to increase of the thermal width for the pairs moving in both parallel and transverse directions. Therefore, we show that the screening phenomena related to the binding energy and the Landau damping phenomena related to the imaginary part of the potential, which both contribute to the dissociation of the Q\={Q} pair, have similar qualitative dependence on the angle and the velocity of the motion. We conclude that the heavy bound states moving in the QGP are dissociated easier than the static ones, while orthogonal motion makes the dissociation easier. Our results have been compared to others coming from the EFT and perturbative QCD and they are in qualitative agreement.

\subsection*{Acknowledgments}
We are thankful to H. Soltanpanahi for useful conversations. The research of D.G. is implemented under the "ARISTEIA" action of the "operational program education and lifelong learning" and is co-funded by the European Social Fund (ESF) and National Resources.


\end{document}